\documentstyle{article}

\newcommand{\beq}{\begin{equation}}
\newcommand{\eeq}{\end{equation}}
\newcommand{\beqa}{\begin{eqnarray}}
\newcommand{\eeqa}{\end{eqnarray}}

\def\<{\langle}
\def\>{\rangle}
\newcommand{\complex}{{\kern .1em {\raise .47ex\hbox {$\scriptscriptstyle |$}}\kern -.4em {\rm C}}}
\newcommand{\real}{{{\rm I} \kern -.19em {\rm R}}}

\title{Bell inequality and the locality loophole: Active versus passive switches}

\author
{N. Gisin and H. Zbinden\\
{\protect\small\em Group of Applied Physics}\\
{\protect\small\em University of Geneva, 1211 Geneva 4, Switzerland}
}
\date{\today}

\begin{document}

\maketitle

\begin{abstract}
All experimental tests of the violation of Bell's inequality suffer from some loopholes.
We show that the locality loophole is not independent of the detection loophole: in
experiments using low efficient detectors, the locality loophole can be closed equivalently 
using active or passive switches.
\end{abstract}


\section{Introduction}\label{int}
Quantum nonlocality, i.e. entanglement of distant systems, plays a central role in 
today's Natural Sciences. It is at the core of quantum physics and its holistic
description of physical systems. These non-local
(but not relativity violating) correlations are nowadays exploited as a resource  for
Quantum Information Processing (QIP). Hence Quantum nonlocality is intimately related to
both relativity and to (Shannon) information theory, bringing together the three major
conceptual scientific breakthroughs of this century.
Some readers may object that information theory is not a Natural Science, but,
precisely, the possibility to exploit entanglement for QIP 
forces us to accept Landauer view that {\it information 
is physical} \cite{Landauer96} and that information theory is a Natural Science 
\cite{FabricReality97}. 

Whatever views one holds on these deep issues, it is clearly desirable to test entanglement,
as it leads to the infamous measurement problem, to the Schr\"odinger cat paradox, to nonlocality
and to quantum information processing. The first two examples just mentioned are unfortunately
not yet directly testable, hence of limited scientific interest (though see the pioneer work
\cite{Haroche96}). QIP
on the other extreme is a very promising scientific field. It has been argued that it
could become the Science of the 21st century, similar to electro-dynamics for the 20th
century \cite{BerryQIP98}. However, almost all the promises of QIP rely on the assumption
that entanglement as described by quantum theory really exists even for systems
composed of tens, hundreds or thousands of subsystems and that entanglement is robust
enough to be manipulated. Unfortunately, in practice, only 2 or 3 particle systems have been
demonstrated so far. Moreover, even for the simplest case of 2 particle entangled systems, all
experimental demonstrations suffer from technological limitations that leave open some
loopholes. Admittedly these loopholes are somewhat artificial. Nevertheless, considering the
conceptual and the practical importance of quantum nonlocality, these loopholes deserve
further investigations.

In this letter, we briefly review the two main loopholes (section \ref{locdetloophole}), 
establish a connection between
them and analyse their status (section \ref{status}) after the recent experiments carried out 
in Innsbruck \cite{BellInnsbruck98} and in Geneva \cite{TittelPRL98,TittelPRA99}.

\section{Bell inequality and the detection and locality loopholes}\label{locdetloophole}
It is not our goal here to present Bell inequality, but simply to fixe the notation. The 
most common form of the Bell inequality reads \cite{BellCHSH}:
\beq
S\equiv E(\alpha,\beta)+E(\alpha,\beta')+E(\alpha',\beta)-E(\alpha',\beta')\le 2
\label{CHSH}
\eeq
where $\alpha$ and $\alpha'$ are two possible settings of a measuring apparatus $A$ analysing
the first subsystem and similarly for $\beta$ and $\beta'$. The measurement outcomes
are labelled $a=\pm1$ and $b=\pm1$. In (\ref{CHSH}), the correlation $E(\alpha,\beta)$ is defined in
function of the coincidence function $C(A=a,B=b|\alpha,\beta)$ as:
\beqa
E(\alpha,\beta)=C(A=1,B=1|\alpha,\beta)+C(A=-1,B=-1|\alpha,\beta)  \nonumber\\
-C(A=1,B=-1|\alpha,\beta)-C(A=-1,B=1|\alpha,\beta)
\eeqa
The assumption of locality, as formulated by Bell in 1964 \cite{Bell64}, 
implies that given the total state
$\lambda$ the result on one side is independent of the setting on the other side:
\beq
P(A=a,B=b|\lambda,\alpha,\beta)=P(A=a|\lambda,\alpha)P(B=b|\lambda,\beta)
\eeq
The
$\lambda$ are called local hidden variables (lhv), although only their local character
is important. The only additional assumption on $\lambda$ is that they belong to some measurable
space (in the mathematical sense), hence that mean values can be computed in the usual way,
with some normalized probability distribution $\rho(\lambda)$:
\beq
C(A=a,B=b|\alpha,\beta)=\int P(A=a|\lambda,\alpha)P(B=b|\lambda,\beta) \rho(\lambda)d\lambda
\eeq

Let us emphasize that determinism is not an issue here. Indeed, the lhv 
$\lambda$ could incorporate enough randomly
chosen data to play any chance game. It is irrelevant whether a coin is tossed at 
the analyzer site or at the source.

The detection loophole \cite{detloophole} 
is based on the fact that in real experiments only a fraction of the
particle pairs emitted by the source are detected. Hence, the sample of detected pairs
could be biased.
A lhv model exploits the detection loophole if the following strict inequality holds:
\beq
P(A=1|\lambda,\alpha)+P(A=-1|\lambda,\alpha) < 1
\label{detloophole}
\eeq
Note that for the detection loophole it is irrelevant whether the particle never gets to
the detector or whether it get there but does not get detected. Actually, such a distinction
is ill defined, first because it has no testable consequences, next because the very concept 
of detector is not sharp enough (is detection of a photon the creation of the first
photon-electron? or the amplification of this? how large must the amplification be? etc). 
For an explicit lhv model based on this loophole reproducing exactly the quantum correlation
see \cite{NGBG99}.

The locality loophole \cite{localityloophole} is based on the fact that in most of the experiments the settings
$\alpha$ and $\beta$ are set long before the particle pairs are produced. Hence it is
logically possible that the source produces lhv with a probability distribution 
$\rho(\lambda)$ which depends on the settings:
\beq
\rho(\lambda)=\rho(\lambda,\alpha,\beta)
\label{locloophole}
\eeq

Both loopholes are based on a similar intuition quite natural from the lhv
point of view. The particle pairs have addition parameters (lhv) that enable them to
answer to certain questions and not to others (ie to pass the analyzer for certain
settings and not for others). If the actual setting does not correspond to the lhv of the
particle, then, according to the detection loophole the particle are simply not detected.
While, according to the locality loophole this situation never happens because the source "knows"
the settings in advance.

In principle both loopholes can be closed by appropriate experiments. To close the detection
loophole, however, one needs detection efficiencies higher than 
$\frac{2}{1+\sqrt{2}}\approx 82.8\%$ \cite{CH74}. No experiment today has achieved this.
Hence, one has to face the annoying fact that the detection loophole resists after
almost 30 years of research and progress!

To close the locality loophole, the settings should be chosen only after the particles 
have left the source, ie after the assumed lhv is fixed. In 1982 A. Aspect and co-workers
were the first to
address this issue \cite{aspectswitch82}. In their remarkable experiment quasi-periodic modulators
with different frequencies selected the settings on both sides (see also the critics
in \cite{ZeilingerAspect86}). More recently experiments in Innsbruck \cite{BellInnsbruck98}
and in Geneva \cite{TittelPRL98,TittelPRA99}
have confirmed the 1982 results, as discussed in the next section.

\section{Active versus passive switches for the locality loophole}\label{status}
In order to examine the locality loophole let us concentrate on Fig. 1. Ideally, the 
settings on Alice and Bob sides should be chosen randomly (by beings enjoying free-will).
The recent beautiful experiment of the Innsbruck group comes very close to this ideal
situation \cite{BellInnsbruck98}. The randomness was provided by a quantum random number generator triggering
a polarization modulator. Moreover
the data were registered locally and compared only in a later stage. In 
Alice's black box as described in Fig.1, the polarization modulator used in the Innsbruck experiment
(i.e. a setting modulator) is replaced by an active switch which
selects the setting $\alpha$ or $\alpha'$ used to analyse the
particle. This is equivalent to the Innsbruck experiment and makes more straightforward
the comparison with the Geneva experiment. The principle of the latter is illustrated in 
Bob's black box of Fig. 1. A beam splitter (in practice a fiber optic coupler) is used as
a passive switch connecting the incoming particle to analysers with setting $\beta$ and
$\beta'$. Only one analyser, either the one with setting $\beta$ or the $\beta'$ one, 
is turned on at a time, while the detectors of the other 
analyser are turned off. Hence the choice of the settings can be done by a purely
electronic switch, as for Alice's box. 

Note that if the active switch in Alice box has 50\% losses, then both
black boxes are undistinguishable. Nevertheless,
It is intuitively clear that the scheme with an active optical switch is closer to the ideal case.
Let us thus analyse Alice box closer, assuming a lossy switch, as in real experiments. One
could argue that the probability distribution of the lhv could still depend on the settings:
the source would randomly guess the setting and whenever the position of the switch is such that
the actual setting differs from the guessed setting, the particle would get lost in the
switch. Such an argument is logically consistent, but admittedly very artificial. Moreover, 
such an argument is closer in spirit to the detection loophole than to the locality loophole.
Accordingly, we argue that an experiment with lossy active switches does close the locality
loophole (assuming that the data violate Bell inequality). Hence, the Innsbruck
experiment does indeed close the locality loophole, although, from pure logic one 
can't exclude (\ref{locloophole}).

The above discussion shows that the two loopholes are not independent and that in practice 
(i.e. for lossy active switches) the implementation with active or passive switches
as illustrated for Alice and Bob in Fig.1 are equivalent. Let us elaborate on the connection
between the two loopholes. A real detector, that is a detector of finite efficiency $\eta$ 
is equivalent to an ideal detector ($\eta_{ideal}=100\%$) with
a passive beamsplitter in front removing a fraction $1-\eta$ of the photons. 
This beam splitter has an open port. It is then natural to
connect another analyzer to this open port. There is clearly no different in 
principle between a detector with a coupler in front when the second output port of the
coupler is left open or when the second port is connected to a turned off analyser. What
is disturbing in this reasoning is that it seems obvious that turning off the analyser
connected to the second port has no influence on the main analyser connected to
the first output port. Moreover, it also seems obvious that the 2 analysers
play a symmetric role. From this discussion we conclude that as long as the active 
switch have losses higher than 50\%, Alice implementation using an active switch is in 
practice as good (or as bad) as Bob's implementation using passive splitter. Thus the Geneva
experiment does also close the locality loophole.

\section{Conclusion}
Quantum nonlocality plays a central role both for our understanding of quantum physics
and for the promising field of quantum information technology. All the experimental
evidence provides an overwhelming support for quantum nonlocality. Nevertheless, some
loopholes prevent the conclusion that lhv are logically excluded. The two main loopholes
have been recalled, defined and some relations established. A first conclusion is that the
use of lossy active switches to test the locality loophole is equivalent to
the use of a passive beam splitter. The second and main conclusion is that the recent
experiments \cite{BellInnsbruck98,TittelPRL98,TittelPRA99} 
confirm Aspect's result and close the locality loophole. The detection
loophole, however, remains embarrassingly open, despite years of efforts.

\small
\section*{Acknowledgments}
Stimulating discussions with Jurgen Brendel, Bruno Huttner, Sandu Popescu, Abner Shimony 
and Woflgang Tittel were appreciated!
This work was partially supported by the Swiss National Science Foundation and by
the European TMR Network "The Physics of Quantum Information" through the Swiss
OFES.

\section*{Figure Captions}
\begin{enumerate}
\item General scheme of tests of Bell's inequality. Alice's (A) and Bob's (B) "black box measurement 
apparatuses" each have two possible settings, a,a' and b,b', respectively, and two possible outcomes r and g.
The inside of both apparatuses are also shown. In Alice's apparatus the setting a or a' determines the 
state of an active optical switch which directs the incoming photons to the corresponding analyser.
In Bob's apparatus the setting b or b' determines that only the detectors of the corresponding analyser
are turned on; hence the passive optical switch (beam splitter) directs the photon at random (in superposition)
but only the selected analyser can detect the photon. If the losses of the active switch are 3 dB (i.e. 50\%) 
higher than the passive switch, then both "black box measurement apparatuses" are undistinguishable.
\end{enumerate}

\end{document}